\documentclass[]{aa} 
\usepackage{times} 
\usepackage{psfig} 
\def\be{\begin{equation}} 
\def\ee{\end{equation}} 
\def\teff{T_{\rm eff}} 
\def\bea{\begin{eqnarray}} 
\def\eea{\end{eqnarray}} 
\def\sss{SSS} 
\def\ssss{SSSs} 
\def\xr{X--ray} 
\def\xs{X--rays} 
\def\rxj{RX\thinspace J0925} 
\def\c87{CAL87} 
\def\sax{BeppoSAX} 
\begin{document}
\thesaurus{08.09.2 \rxj, 08.06.3, 08.01.3, 13.25.5, 08.23.1}
\title{Constraining the spectral parameters of RX\thinspace J0925.7--4758 with 
the \sax\ LECS} 
\author{H.W.~Hartmann$^1$ \and J.~Heise$^1$ \and P.~Kahabka$^2$ \and 
C.~Motch$^3$ \and A.N.~Parmar$^4$}
\authorrunning{H.W. Hartmann et al.}
\institute{$^1$SRON Laboratory for Space Research, Sorbonnelaan 2, 3584 CA 
Utrecht, the Netherlands\\
$^2$Astronomical Institute `Anton Pannekoek', University of Amsterdam, 
Kruislaan 403, 1098 SJ Amsterdam, the Netherlands\\
$^3$Observatoire Astronomique, UMR 7550 CNRS, 11 rue de l'Universit\'e, F-67000 
Strasbourg, France\\
$^4$Astrophysics Division, Space Science Department of ESA, ESTEC, P.O. Box 
299, 2200 AG, Noordwijk, the Netherlands} 
\offprints{w.hartmann@sron.nl} 
\date{Received ; accepted } 
\maketitle 
\begin{abstract} 
The Super Soft Source RX\thinspace J0925.7--4758 (\rxj\ hereafter) was observed 
by \sax\ LECS and MECS on January 25--26 1997. The source was clearly detected 
by the LECS but only marginally detected by the MECS. We apply detailed 
Non-Local Thermodynamic Equilibrium (Non-LTE) models including metal line 
opacities to the observed LECS spectrum. We test whether the \xr\ spectrum of 
\rxj\ is consistent with that of a white dwarf and put constraints upon the 
effective temperature and surface gravity by considering the presence or 
absence of spectral features such as absorption edges and line blends in the 
models and the observed spectrum.
\par
We find that models with effective temperatures above $\sim10^6\mbox{ K}$ or 
below $\sim7.5\times 10^5\mbox{ K}$ can be excluded. If we assume a single 
model component for \rxj\ we observe a significant discrepancy between the 
model and the data above the Ne\,{\sc ix} edge energy at 1.19 keV. This is 
consistent with earlier observations with ROSAT and ASCA. The only way to 
account for the emission above $\sim1.2\mbox{ keV}$ is by introducing a second 
spectral (plasma) component. This plasma component may be explained by a 
shocked wind originating from the compact object or from the irradiated 
companion star.
\par
If we assume $\log g = 9$ then the derived luminosity is consistent with that 
of a nuclear burning white dwarf at a distance of $\sim 4$ kpc.
\keywords{Stars: individual (RX\thinspace J0925.7--4758) - Stars: fundamental 
parameters - Stars: atmospheres - X-rays: stars - White dwarfs}
\end{abstract} 
\section{Introduction} 
\subsection{Supersoft \xr\ Sources} 
\label{sss} 
Observed with low spectral resolution satellites like Einstein and ROSAT, 
Supersoft X-ray Sources (SSSs) show heavily absorbed featureless spectra. They 
are generally characterized by the ROSAT hardness ratio of
\bea
{\rm HR1} \equiv {{\rm H - S} \over {\rm H + S}} \leq -0.8 ,
\nonumber
\eea
where H and S are the count rates between 0.5--2.0 keV and 0.1--0.4 keV 
respectively. This implies that the bulk of the observed photons have energies 
below 0.5 keV. When fitted with blackbody spectra \sss\ temperatures typically 
lie in the range $1.5-5\times 10^5\mbox{ K}$. The derived column densities are 
of the order of $n_{\rm H}=10^{20}-10^{22}\mbox{ cm}^{-2}$. Many of the \ssss\ 
have now been optically identified, covering a range of different objects 
including e.g. classical novae, symbiotic novae and the central star of the 
planetary nebula N67. Thus \ssss\ probably do not form a homogeneous class but 
they may all contain a hot white dwarf which emits the observed soft \xs. They 
are commonly believed to be nuclear burning white dwarfs. The nuclear fuel is 
supplied by a main sequence (or a slightly evolved) companion star or, in the 
case of N67, by primordial hydrogen and helium left after the asymptotic giant 
branch phase. This accretion takes place in a very narrow range of rates 
between $3 \times 10^{-8}$ and $7 \times 10^{-7}{\rm 
M}_{\odot}\:\mbox{yr}^{-1}$. For the steady burning \ssss\ this accretion rate 
is critical: at lower accretion rates white dwarfs show nova-like behavior. At 
higher rates white dwarfs develop a strong wind that compensates for the 
accretion rate (Paczy\'nski \& $\dot{\rm Z}$ytkow \cite{pazy}; Sion et al. 
\cite{sion}; Sienkewicz \cite{sien}; Hachisu et al. \cite{hachisu}). 
\par
\begin{figure}[t] 
\psfig{file=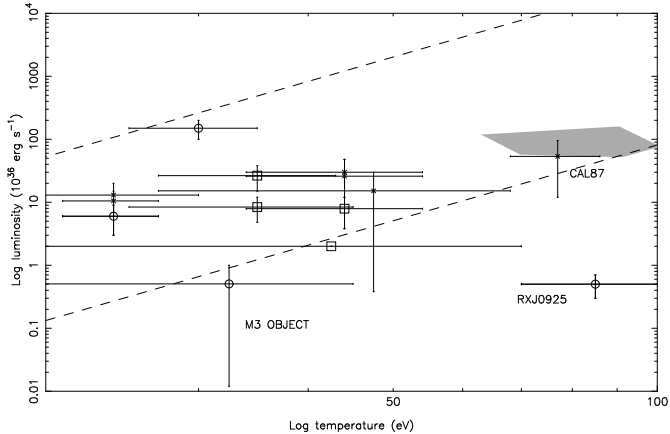,width=8.8cm} 
\caption{Luminosity versus temperature for several \ssss. Crosses: LMC 
sources. Squares: SMC sources. Circles: Galactic sources. The dashed lines 
indicate the relation $L_{\rm bol}=4\pi R^2\:\sigma T^4$ for $R\approx2.5\times 
10^8\mbox{ cm}$ (lower line) for the most massive white dwarfs and $5 \times 
10^9\mbox{ cm}$. A distance of 1 kpc has been assumed for \rxj. Data are 
taken from Kahabka \& Van den Heuvel (1997). For the explanation of the shaded 
area we refer to the text in Sect. \ref{hot_model}} 
\label{lt_plot} 
\end{figure}

 
CAL87 and RX\thinspace J0925.7--4758 (\rxj\ hereafter) are the only two sources 
which have substantial count rates above 0.5 keV up to $\sim1.0\mbox{ keV}$. 
Those sources are fitted with relatively high temperatures of $\la 7.5\times 
10^5\mbox{ K}$ and $\la 9\times 10^5\mbox{ K}$ respectively. Fig. \ref{lt_plot} 
shows the luminosity versus temperature for several \ssss\ with known 
temperature and luminosity. Data are taken from Kahabka \& Van den Heuvel 
(\cite{kah_heu}). Also indicated are two dashed lines which represent the 
luminosity at $2.5\times 10^8\mbox{ and } 5 \times 10^9\mbox{ cm}$ radius 
respectively in the relation $L_{\rm bol}=4\pi R^2\:\sigma T^4$. This range in 
radius applies to white dwarfs. Note that most \ssss\ lie within this range. 
The lower limit for the radius corresponds to the Chandrasekhar upper mass 
limit for white dwarfs of $1.4 M_{\sun}$, applying the mass-radius relation 
derived by Kahabka \& Portegies Zwart (\cite{spz}). \rxj\ is shown in the same 
figure assuming a distance to the source of 1 kpc. Temperature and luminosity 
bounds have been derived by Hartmann \& Heise (\cite{hartmann}) by fitting 
model atmosphere spectra to ROSAT PSPC data of \rxj. Though a good fit was 
obtained ($\chi^2=0.78$) the derived normalization of the model which is 
defined as $R^2/D^2$ is obviously too low, $R$ is the radius of the compact 
object and $D$ is 1 kpc. A small radius for the compact object (less than $\sim 
1000$ km) poses a problem for the white dwarf hypothesis. It has been suggested 
that \rxj\ is at a distance of 10 kpc in order to account for the factor 
$\sim100$ needed to increase the luminosity into the regime bounded by the two 
dashed lines in Fig. \ref{lt_plot} and thus to obtain a model for \rxj\ 
consistent with a white dwarf (Ebisawa et al. \cite{ebis2}; Motch 
\cite{motch3}).
\par 
In order to be able to constrain the model parameters more tightly than was 
previously possible with the ROSAT PSPC and to test whether \rxj\ can be 
consistent with a white dwarf, \rxj\ was observed with \sax. We have refined 
our previous model spectra fitted to \rxj\ by including line opacities in the 
model atmospheres. In this Paper we describe the data analysis of this 
observation.
\subsection{\rxj} 
\label{rxj} 
The Supersoft \xr\ Source \rxj\ was discovered in the ROSAT Galactic Plane 
Survey (RGPS, Motch et al. \cite{motch1}). This survey is defined as the 
$\mid\! b\!\mid \leq 20\degr$ region of the ROSAT All Sky Survey. By fitting a 
blackbody to the ROSAT observation a column density of $n_{\rm H}=(1.4-3.7) 
\times 10^{22}\mbox{ cm}^{-2}$ was derived. Optical observations indicate 
$n_{\rm H}=(1.0-1.9) \times 10^{22}\mbox{ cm}^{-2}$. This relatively high 
column density indicates that the source is located within or behind the nearby 
Vela Sheet molecular cloud (Motch et al. \cite{motch2}). This puts \rxj\ 
at a distance of at least 425 pc. The galactic coordinates are 
$\mbox{l\sc ii}=271\fdg36\mbox{ and b\sc ii}=1\fdg88$. Considering this 
low latitude one can conclude that \rxj\ is a galactic source. A transient jet 
with a peak jet velocity of $5200\mbox{ km s}^{-1}$ was detected from optical 
observations. This is compatible with the escape velocity of both a massive 
white dwarf (Motch \cite{motch3}) and a shrouded neutron star with an extended 
photosphere of a few thousand kilometers (Greiner \cite{greiner91}).
\par 
Ebisawa et al. (\cite{ebis1}) observed \rxj\ with the ASCA Solid-state Imaging 
Spectrometer (SIS) in December 1994. A single blackbody spectrum does not fit 
their data. Applying additional absorption edges at $\sim0.9$, 1.0 and 1.4 keV 
and reducing the abundances of interstellar oxygen and iron to A(O) = 0.38 and 
A(Fe) = 0.2 times solar result in an acceptable fit. The edges at $\sim0.9$ and 
1.4 keV suggest the presence of atmospheric O\,{\sc viii} and Ne\,{\sc x} or 
Fe\,{\sc xviii}. However, the O\,{\sc viii} edge is coincident with the 
interstellar neon absorption edge and its identification is therefore not 
unambiguous. The origin of a possible edge at $\sim1.0\mbox{ keV}$ remains 
unclear and the observed spectral shape around $\sim 1\mbox{ keV}$ may be due 
to strong line opacities as well. Preliminary fits of Non-Local Thermodynamic 
Equilibrium (Non-LTE) model spectra at $\log g=9.0\mbox{ cm s}^{-2}$ to 
ASCA--SIS data show a strong Ne\,{\sc ix} edge in the model which is not 
observed in the data and a small normalization of 160 km (D/kpc). The fitted 
effective temperature and column density are $\sim8.5\times 10^5\mbox{ K and } 
1.0 \times 10^{22}\mbox{ cm}^{-2}$ respectively (Ebisawa et al. \cite{ebis1}). 
These parameters are consistent with those derived by Hartmann \& Heise 
(\cite{hartmann}) from the ROSAT PSPC observation. 
\section{\sax\ observation of \rxj} 
\label{obs} 
\rxj\ was observed on January 25--26 1997 with the Low and Medium Energy 
Concentrator Spectrometer (LECS and MECS respectively) aboard \sax. The LECS 
energy resolution is a factor $\sim2.4$ better than that of the ROSAT PSPC, 
while the effective area is about a factor $\sim20\mbox{ to }5$ lower at 
0.28 and 1.5 keV respectively. Since the LECS can only be operated during 
satellite night-time, the LECS observed \rxj\ for only 44 ksec. \rxj\ is 
clearly detected with the LECS as an on-axis source with an average net 
count rate of $0.074 \pm 0.002\mbox{ cts s}^{-1}$. 
\par 
We rebinned the spectral data to ${1 \over 3} \times$ FWHM of the spectral 
resolution quoted by Parmar et al. (\cite{parmar2}) and Boella et al. 
(\cite{boella}). Moreover, we require a minimum of 20 counts per energy bin to 
allow the use of the $\chi^2$ statistic. The resulting 18 energy bins are used 
for spectral analysis. The standard on-axis 35 pixel radius region, which is 
supplied together with the \sax\ data analysis package, is used for extracting 
source photons. Background photons are subtracted applying the same 35 pixel 
region to a \sax\ deep field exposure. During the 103 ksec. exposure with the 
MECS \rxj\ is only marginally detected on-axis in the 1.3 -- 2.0 keV energy 
band with an average count rate of $0.0022\pm 0.0006\mbox{ cts s}^{-1}$. Above 
2.0 keV \rxj\ is not detected with the MECS. In this paper we will therefore 
focus our attention upon the LECS data. 
\section{White dwarf model spectra} 
\label{models} 
\begin{table*}[t] 
\caption[]{Overview of the model atoms. Line opacities are included only for H, 
He, C, N, O, Ne and Fe. The lines in the model are Non-LTE. The lines in the 
spectrum are selected from an extensive line list. The ratio of the line center 
opacity to the continuum opacity should be at least $10^{-4}$ for the line to 
be selected. A selected line that is not in the model atmosphere is treated in 
LTE.\\$^*$: The exact number of lines for each ion in the model spectrum varies 
with model parameters like e.g. the effective temperature}
\label{mod_tab}
\begin{flushleft} 
\begin{tabular}{llllllll} 
\noalign{\smallskip} 
\hline\noalign{\smallskip} 
Ion & No. levels & No. lines & No. lines$^*$ & Ion & No. levels & No. lines & 
No. lines$^*$ \\ 
 & & (model) & (spectrum) & & & (model) & (spectrum) \\ 
\noalign{\medskip} 
\hline\noalign{\smallskip} 
H\,{\sc i}    & 5  & 10 & --      & Ne\,{\sc x}     & 10 & 45  & $\ga10$   \\
He\,{\sc ii}  & 3  & 3  & --      & Fe\,{\sc xvi}   & 15 & 44  & $\ga70$   \\
C\,{\sc vi}   & 10 & 45 & $\ga20$ & Fe\,{\sc xvii}  & 21 & 43  & $\ga1000$ \\
N\,{\sc vii}  & 10 & 45 & $\ga30$ & Fe\,{\sc xviii} & 50 & 256 & $\ga1000$ \\
O\,{\sc vii}  & 17 & 33 & $\ga50$ & Fe\,{\sc xix}   & 21 & 31  & $\ga1000$ \\
O\,{\sc viii} & 10 & 45 & $\ga30$ & Fe\,{\sc xx}    & 21 & 41  & $\ga1000$ \\
Ne\,{\sc viii}& 1  & 1  & $\ga20$ & Fe\,{\sc xxi}   & 21 & 36  & $\ga1000$ \\
Ne\,{\sc ix}  & 17 & 33 & $\ga70$ & Fe\,{\sc xxii}  & 21 & 52  & $\ga1000$ \\
\noalign{\smallskip} 
\hline 
\end{tabular} 
\end{flushleft} 
\end{table*} 
In the \xr\ regime optically thick spectra are often still fitted with 
continuum models like e.g. blackbodies, sometimes combined with power law or 
bremsstrahlung spectra. Due to the moderate spectral resolution of present day 
\xr\ detectors operating in space individual lines are not resolved and only 
line blends (if present in the spectrum) can be distinguished. However, it has 
been argued that optically thick model spectra that only involve continuum 
opacities do not represent the observed spectrum correctly. The effect of many 
line opacities (line blanketing) is to change the temperature structure of the 
model atmosphere and therefore the ionization balance and continuum shape of 
the model spectrum. 
\par 
Moreover, Hartmann \& Heise (\cite{hartmann}) note that the appearance of 
emission edges in Non-LTE model spectra may be due to the absence of strong 
opacity sources other than that of the ionization edge. By including line 
opacities in the model the emission edges may either be reduced in strength, 
disappear or even turn into absorption edges. 
\par 
We will fit the \sax\ spectrum of \rxj\ with detailed Non-LTE model spectra, 
see Table \ref{mod_tab}. In order to test wether \rxj\ can be consistent with a 
white dwarf the applied range in surface gravity is $7.5 \le \log g \le 9.5 
\mbox{ cm s}^{-2}$. Non-LTE models differ from LTE models in the sense that the 
atomic level populations are allowed to deviate from the Saha-Boltzmann 
distribution. Nowadays it takes little extra time to calculate more 
sophisticated Non-LTE atmosphere models. However, it has been shown that hot 
high-gravity Non-LTE spectra can be significantly different from LTE model 
spectra (e.g. Hartmann \& Heise \cite{hartmann}). For the calculation of the 
model atmospheres we have used the atmosphere code TLUSTY and for the 
calculation of the model spectra we have used the code SYNSPEC (Huben\'y 
\cite{hubeny}, Huben\'y \& Lanz \cite{hula}). Solar abundances are according to 
Anders \& Grevesse (\cite{anders}). Photoionization cross-sections of the 
ground states are calculated using data and fitting formula from Verner \& 
Yakovlev (\cite{verner}). Photoionization cross-sections for higher levels, 
energy levels and oscillator strengths are taken from the Opacity Project 
database (Cunto et al. \cite{cunto}). The line profiles for the model 
calculations are approximated by a Doppler profile ignoring turbulent 
velocities. Voigt profiles which take into account the effects of natural, 
Stark and thermal Doppler broadening have been used for the calculation of 
spectral lines. The collisional rates are given by (${\rm i}<{\rm j}$): 
\bea
C_{\rm ij} = n_{\rm e}\Omega_{\rm ij}(T)
\nonumber
\eea
For hydrogen and helium, $\Omega_{\rm ij}(T)$ is taken from Mihalas et al. 
(\cite{mha}). For other elements Seaton's equation for collisional ionization 
(\cite{seat}) and van Regemorter's equation for collisional excitation 
(\cite{regem}) with $\bar{g}=0.25$ are used. Due to the computing times 
required to calculate a detailed model atmosphere we will apply only small 
grids of Non-LTE spectra that contain detailed model atoms including line 
opacities of H, He, C, N, O, Ne and Fe cf. Table \ref{mod_tab}.
\section{\sax\ data of \rxj} 
\subsection{Spectral analysis of \rxj} 
\begin{table*} 
\caption[]{Fit results to the \sax\ observation of \rxj. The radii are obtained 
from the normalization area assuming spherically symmetric emission. The errors 
quoted are the 90\% confidence intervals.\\
GRID1 is similar to the H--Ne models applied in Hartmann \& Heise 
(\cite{hartmann}) and Ebisawa et al. (\cite{ebis1}), but now with iron as well. 
The models include continuum opacities only. GRID2 cf. GRID1 but now including 
line opacities for all elements H, He, C, N, O, Ne and Fe. GRID3 contains 
relatively hot models, i.e. the effective temperatures go up to $1.2\times 
10^6\mbox{ K}$ and $\log g$ is fixed at 9.5\\
$^*$: Fixed parameter}
\label{fitres}
\begin{flushleft} 
\begin{tabular}{ccccccc} 
\noalign{\smallskip} 
\hline 
\noalign{\smallskip} 
 & T & $\log\mbox{ g }$ & $n_{\rm H}$ & L & R & $\chi^2_{\rm red.}$ (dof) \\ 
 & $10^5\mbox{ K}$ & $\mbox{cm s}^{-2}$ & $10^{21}\mbox{ cm}^{-2}$ & $10^{35} 
\mbox{erg s}^{-1}\mbox{(d/kpc)}^2$ & $10^7\mbox{ cm (d/kpc)}$ & \\ 
\noalign{\smallskip} 
\hline\noalign{\smallskip} 
GRID1 & $8.71_{-0.07}^{+0.07}$ & $9.0^*$ & $ 8.2_{-0.8}^{+1.0}$ 
& $1.1_{-0.4}^{+0.7}$    & $ 1.6_{-0.3}^{+0.5}$   &  6.0 (15) \\ 
\noalign{\smallskip} 
GRID2 & $8.75_{-0.03}^{+0.05}$ & $9.0^*$                & $ 15.1_{-1.7}^{+1.8}$ 
& $41_{-22}^{+51}$      & $ 10_{-3}^{+6}$   &  5.6 (15) \\ 
\noalign{\smallskip} 
GRID3 & $10.63_{-0.03}^{+0.03}$ & $9.5^*$                & $ 7.6_{-0.4}^{+0.4}$ 
& $0.79_{-0.14}^{+0.14}$ & $0.98_{-0.08}^{+0.08}$ & 11   (15) \\ 
\noalign{\medskip} 
\hline 
\end{tabular} 
\end{flushleft} 
\end{table*} 
The spectral analysis was performed using the SPEX software package (Kaastra et 
al. \cite{kaas}). In previous modeling of ROSAT data (Hartmann \& Heise 
\cite{hartmann}) and ASCA data (Ebisawa et al. \cite{ebis1}) \rxj\ has been 
fitted to Non-LTE model spectra that only involve continuum opacities at 
effective temperatures above $\sim7.5\times 10^5\mbox{ K}$. Based upon this 
effective temperature we expect the surface gravity to be $8.5\la\log g\la 
9.5\mbox{ cm s}^{-2}$. At lower gravity \rxj\ radiates at super-Eddington 
luminosity. At higher gravity the corresponding mass of the assumed white dwarf 
exceeds the Chandrasekhar mass limit (cf. the mass-radius relation for white 
dwarfs Kahabka \& Portegies Zwart \cite{spz}). 
\par 
Our first step is to fit the \sax\ LECS data of \rxj\ to Non-LTE model spectra 
(GRID1) similar to those that have been used to fit the ASCA and ROSAT data in 
order to check for obvious changes since the ASCA observation in 1994. Line 
opacities are omitted from those models which include continuum opacities only. 
We have assumed a value of $\log g=9.0$. The model spectra are folded with 
interstellar absorption using abundances according to Anders \& Grevesse 
(\cite{anders}). Though the fit is not acceptable (reduced $\chi^2=6.0$), the 
parameters are consistent with those obtained by Hartmann \& Heise 
(\cite{hartmann}) and Ebisawa et al. (\cite{ebis1}), see Table \ref{fitres}, 
GRID1. 
\par
\begin{figure}[t] 
\psfig{file=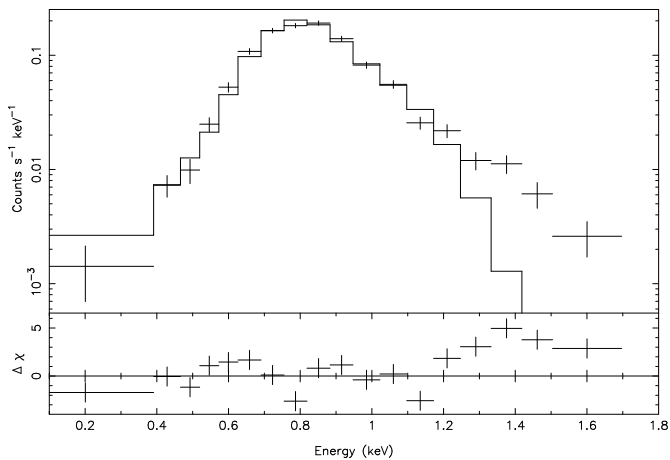,width=8.8cm} 
\caption{Non-LTE model spectrum fit (GRID1) to the \sax\ data of \rxj. Solar 
abundances are assumed. Note the excess flux in the data above 1.2 keV} 
\label{cosm} 
\end{figure}

\begin{figure}[t] 
\psfig{file=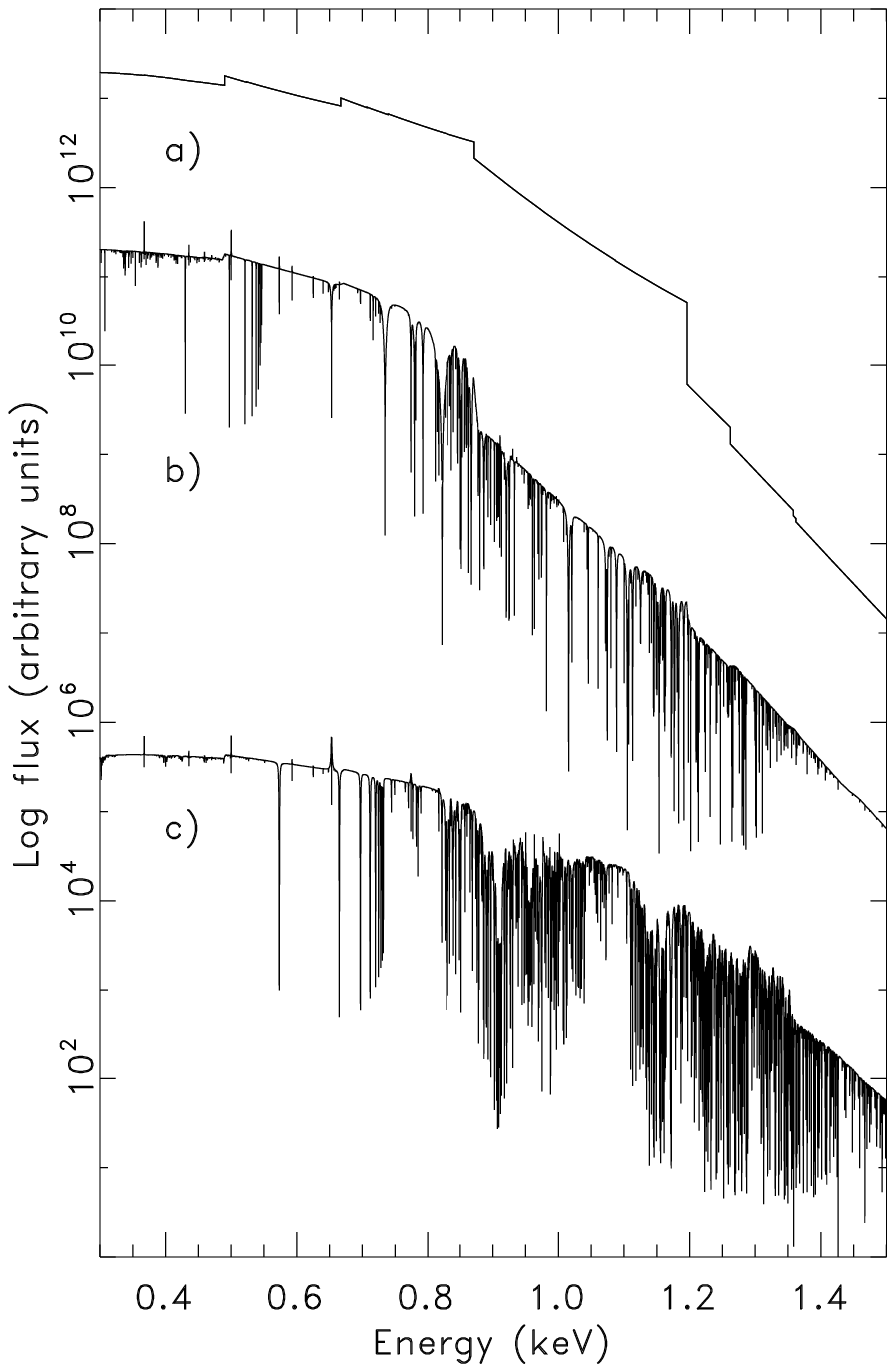,width=8.8cm} 
\caption{({\bf a}) The best-fit GRID1 model spectrum, the spectrum folded with 
the detector response is shown in Fig. \ref{cosm}. The cutoff at the Ne\,{\sc 
ix} edge at 1.19 keV in the model is a likely cause for the discrepancy between 
the model and the data. ({\bf b}) The best-fit GRID2 model spectrum. The model 
atoms for H, He, C, N, O, Ne and Fe are treated in more detail including line 
opacities. Note the increased strength of the O\,{\sc viii} edge at 0.87 keV 
and the reduced strength of the Ne\,{\sc ix} edge at 1.2 keV. ({\bf c}) The 
best-fit GRID3 model spectrum. Iron line blends decrease the flux between 
0.8--1.0 keV and above 1.1 keV}
\label{model} 
\end{figure}

 
Fig. \ref{cosm} shows that the data is in excess of the model above $\sim
1.2\mbox{ keV}$. In Fig. \ref{model}a we have drawn the fitted model spectrum. 
This graph demonstrates that the excess in the data is due to the presence of 
Ne\,{\sc ix} and Fe\,{\sc xvii} absorption edges in the model at 1.19 and 1.26 
keV respectively. This is consistent with the work done by Ebisawa et al. 
(\cite{ebis2}) in their analysis of the ASCA data. They conclude that there is 
only weak evidence for the presence of the O\,{\sc viii} edge at 0.87 keV and 
the Ne\,{\sc ix} edge at 1.2 keV. However, the ASCA data allow for the presence 
of the Ne\,{\sc x} edge at 1.36 keV. 
\par
We demonstrate the effects of adding line opacities to the formerly fitted 
model atmosphere of \rxj. We have calculated a small model grid (GRID2) of 
Non-LTE model spectra and included line opacities of H, He, C, N, O, Ne and Fe, 
see Table \ref{mod_tab}. Using the best-fit temperature (which is consistent 
with the temperature found by applying GRID1), this spectrum is shown as well 
in Fig. \ref{model}b. From this graph it becomes clear that the part of the 
spectrum that is of special interest to us, from 0.5 to 1.5 keV, changes 
significantly when spectral lines are included. Two changes in particular have 
important consequences for the fit to \rxj: 
\begin{enumerate} 
\item{The Ne\,{\sc ix} edge at 1.2 keV has almost completely disappeared in the 
spectrum, although it is included in the model atmosphere. This is expected to 
improve the fit to the excess in the data above 1.2 keV.} 
\item{Instead the O\,{\sc viii} edge at 0.87 keV has become stronger. This 
strong O\,{\sc viii} edge will decrease the model flux between 0.9--1.2 keV.} 
\end{enumerate} 
Line transitions affect the level populations and a new equilibrium between the 
different ionization stages is obtained. In this case the O\,{\sc viii} edge 
has become more prominent (which is not consistent with the findings by Ebisawa 
et al. (\cite{ebis2})) and the Ne\,{\sc ix} edge has completely disappeared. 
Applying GRID2 to the data results in $\chi^2=5.6$. Thus sophisticating the 
originally suggested model does not result in an acceptable fit, though the 
derived absorption column density is now consistent with the value derived from 
ROSAT and optical observations reported by Motch et al. (\cite{motch2}) and the 
luminosity has increased by a factor of $\sim 40$.
\par 
We proceed with comparing the strengths of line blends (since individual lines 
are not resolved by the \sax\ LECS) and absorption edges in the model spectra 
with the observed spectrum. The main aspects under consideration are: 
\begin{enumerate} 
\item{Strong absorption edges in model spectra that are not observed in the 
data, like that of the Ne\,{\sc ix} edge, may be decreased by invoking a high 
ionization stage for that particular atom. Comparing Figs. \ref{cosm} and 
\ref{model}a,b and considering the results of Ebisawa et al. (\cite{ebis2}) 
we expect only a shallow O\,{\sc viii} edge.} 
\item{The observed spectrum may consist of two (or more) components. The 
dominant component is the white dwarf model spectrum. An excess at the 
high-energy end of the data may be explained by a second component of yet 
unknown origin.} 
\end{enumerate} 
In the sections that follow we will study the effects of these two 
possibilities upon the spectrum of \rxj. Note that the first aspect affects 
the structure of the model atmosphere and may therefore reflect upon the model 
spectrum as a whole.
\subsection{Case 1: A highly ionized model atmosphere} 
\label{hot_model} 
\begin{figure}[t] 
\psfig{file=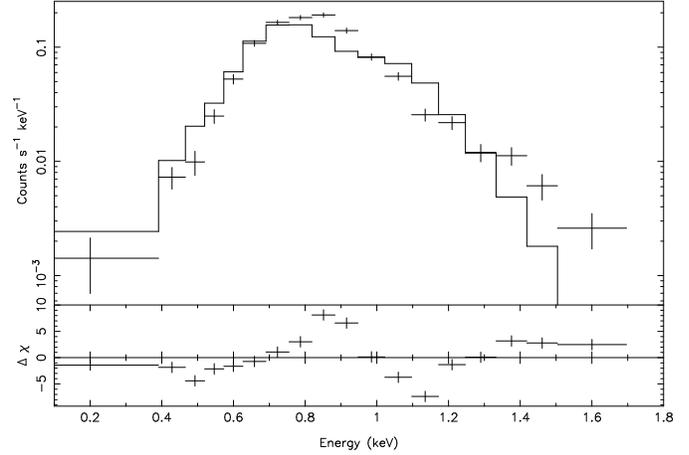,width=8.8cm} 
\caption{The \sax\ spectrum of \rxj\ fitted to the hot ($> 10^6\mbox{ K}$) 
model spectra of GRID3. Strong iron line blends in the range 0.8--1.3 keV cause 
the model to diverge from the data} 
\label{feblend} 
\end{figure} 
At high effective temperatures which are close to the Eddington limit, 
atmospheric neon and iron will become ionized up to Ne\,{\sc x} and beyond 
Fe\,{\sc xviii} respectively. In that case the Ne\,{\sc ix} and Fe\,{\sc xvii} 
edges are shallow or not present at all. Though the hottest models can be 
achieved with the highest reasonable surface gravity $\log g=9.5$, we must keep 
in mind that increasing the surface gravity will counteract the effects of 
increasing the effective temperature with respect to the ionization balance 
(cf. Hartmann \& Heise \cite{hartmann}). The ionization balance is shifted 
towards the lower ionization stages because of the increased electron density 
and recombination rate at higher surface gravity. Our goal is to reduce the 
strength of the O\,{\sc viii} edge and minimize the Ne\,{\sc ix} edge by 
exploring the $\teff, \log g$ parameter space. An additional problem is caused 
by the interstellar Ne\,{\sc i} absorption edge which coincides with the 
atmospheric O\,{\sc viii} edge. For this reason it is difficult to determine 
the strength of the O\,{\sc viii} edge, although the interstellar absorption 
strongly affects the entire lower energy part of the \xr\ spectrum and is 
therefore better confined. 
\par 
We have calculated a small grid of Non-LTE model spectra in conformance with  
Table \ref{mod_tab} in the effective temperature range $10^6\leq\teff\leq 
1.2\times 10^6\mbox{ K}$ and $\log g=9.5$ (GRID3). At effective temperatures 
close to the Eddington limit the model spectrum shows strong Fe\,{\sc xviii} 
and Fe\,{\sc xix} line blends roughly between 0.8--1.1 keV and above 1.3 keV. 
See Fig. \ref{model}c. When folded through the detector response the iron line 
blends cause the model spectrum to become shallower than the data (see Fig. 
\ref{feblend}) resulting in an extremely poor fit. Lowering the effective 
temperature immediately results in an increase in the Ne\,{\sc ix} edge depth. 
We therefore reject extremely hot models that include iron as a model for the 
spectrum of \rxj. The validity of assuming iron-depleted model atmospheres is 
discussed in Sect. \ref{1comp}. Note from Table \ref{fitres} that applying this 
model spectrum with a high effective temperature result in an even lower 
normalization (luminosity). 
\par 
\begin{figure}[t]
\psfig{file=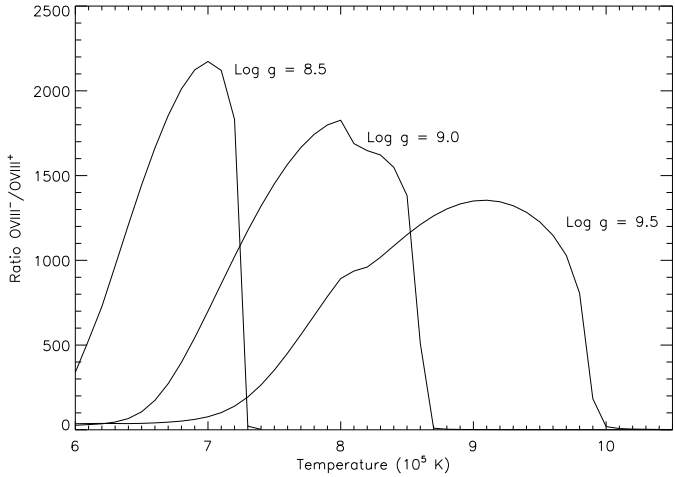,width=8.8cm}
\caption{Steepness of the O\,{\sc viii} transition indicated by the ratio of 
the continuum flux before (O\,{\sc viii}$^-$) and after (O\,{\sc viii}$^+$) the 
transition. For each value of the surface gravity there is an effective 
temperature at which the depth of the O\,{\sc viii} edge decreases rapidly and 
oxygen becomes completely ionized. The region where the O\,{\sc viii} edge is a 
strong spectral feature corresponds to region below the shaded area in Fig. 
\ref{pspace}}
\label{jump}
\end{figure}
\begin{table*}[t]
\caption[]{Fit results combining GRID2 with a Collisional Ionization 
Equilibrium (CIE) plasma cf. Kaastra et al. (\cite{kaas}). The (electron) 
temperature of the plasma is held fixed between 0.2 and 1 keV during the fits
keV.\\ 
$^*$: Fixed parameter}
\label{2ndcomp}
\begin{flushleft} 
\begin{tabular}{cccccccc} 
\noalign{\smallskip} 
\hline 
\noalign{\smallskip} 
 & T & $\log\mbox{ g }$ & $N_{\rm h}$ & L & R & ${\rm N_eN_hV}$ & $\chi^2_{\rm 
red.}$ (dof) \\ 
 & $10^5\mbox{ K}$ & $\mbox{cm s}^{-2}$ & $10^{21}\mbox{ cm}^{-2}$ & $10^{35} 
\mbox{erg s}^{-1}\mbox{(d/kpc)}^2$ & $10^7\mbox{ cm (d/kpc)}$ & 
$10^{56}\mbox{ cm}^{-3}$ & \\ 
\noalign{\smallskip} 
\hline\noalign{\smallskip} 
GRID2 & $ 8.75_{-0.03}^{+0.04}$ & $9.0^*$ & $ 14.2_{-1.5}^{+1.4}$ & $ 
29_{-16}^{+18}$ & $ 9_{-2}^{+3}$ & -- & \\
\noalign{\vspace{-1.5mm}}
 + & & & & & & & 3.4 (14) \\
\noalign{\vspace{-1mm}}
CIE     & $140-17$ & -- & -- & 0.0044-1.5 & -- & 
0.15--60. & \\ 
\noalign{\medskip} 
\hline 
\end{tabular} 
\end{flushleft} 
\end{table*}
Ebisawa et al. (\cite{ebis2}) conclude from the analysis of the ASCA data that 
there is only weak evidence for the presence of the O\,{\sc viii} edge at 0.87 
keV. Therefore, we calculate a model grid including the elements H, He, C, N, O 
and Ne for the surface gravity and effective temperature regimes $8.5\leq\log g 
\leq 9.5$ (though we no longer consider $\log g\approx 8.5$ an appropriate 
option for \rxj) and $6\times 10^5\leq\teff$ up to the respective Eddington 
limit. The purpose is to get an indication of the combination of the effective 
temperature and surface gravity for which oxygen becomes completely ionized and 
the O\,{\sc viii} edge disappears. Iron is omitted from these models since the 
combination $\teff,\log g$ at which O\,{\sc viii} ionizes is hardly affected by 
it and considerable computing time is saved in this way. The ratio of the 
continuum flux just before and after the transition energy of 0.87 keV is 
plotted against the effective temperature in Fig. \ref{jump}. This shows for 
three values of the surface gravity that the transition towards completely 
ionized oxygen by increasing the effective temperature is relatively sharp.
\par
\begin{figure}[t] 
\psfig{file=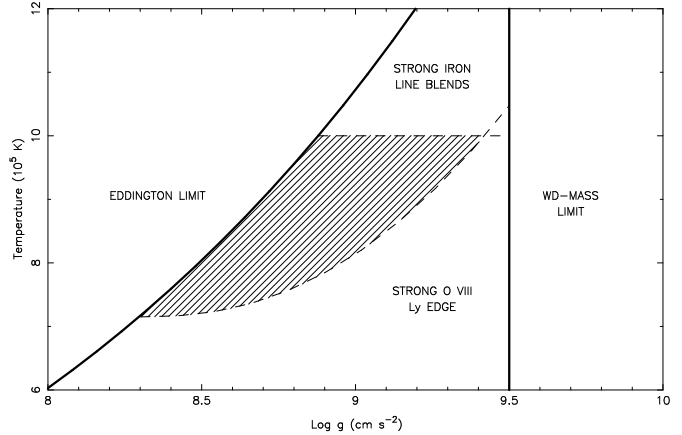,width=8.8cm} 
\caption{$\log g,\teff$ parameter space. The shaded area indicates the region 
applicable to \rxj. The area is bounded by the Eddington stability limit, the 
Chandrasekhar mass limit for white dwarfs and the regions for which a strong 
O\,{\sc viii} edge or the iron line blends should be observed} 
\label{pspace} 
\end{figure}

 
We can now constrain the $\teff,\log g$ parameter space for \rxj\ which is 
indicated in Fig. \ref{pspace}. It is bounded by the Eddington limit, the 
Chandrasekhar limit for white dwarfs (using the mass-radius relation cf. 
Kahabka \& Portegies Zwart (\cite{spz})) and roughly by the temperatures for 
which the O\,{\sc viii} edge and the iron line blends become pronounced (but 
unobserved) features. Note that relatively `cool' models ($\teff\la 
7.5\times 10^5\mbox{ K}$) are ruled out because of the non-detection of a 
strong O\,{\sc viii} edge. We have indicated the corresponding region in Fig. 
\ref{lt_plot}. 
\subsection{Case 2: A second spectral component} 
\label{comp_solution}
\begin{figure}[t] 
\psfig{file=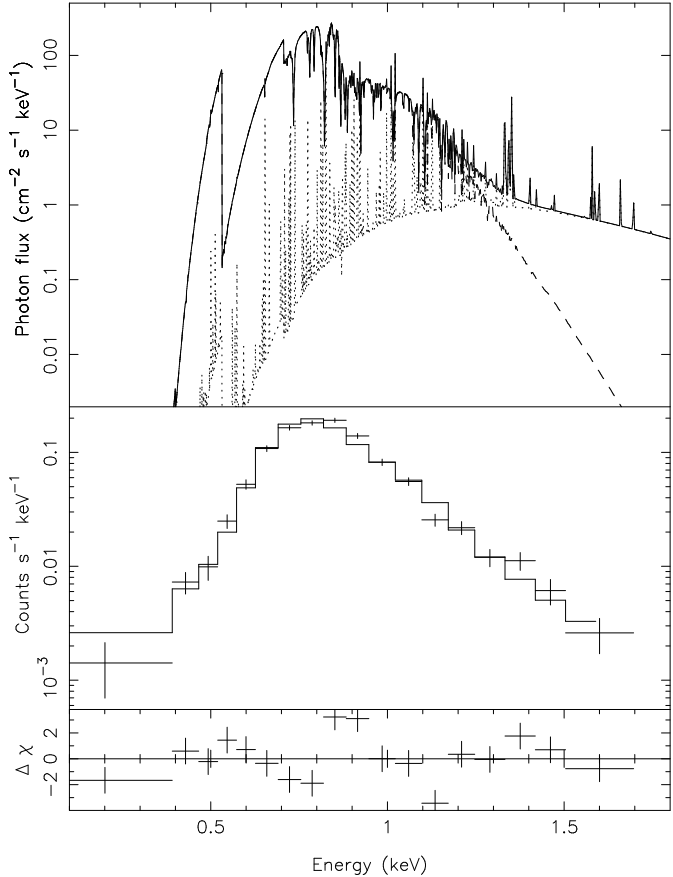,width=8.8cm} 
\caption{Top panel: Photon spectrum of the best fit two-component model 
including the column absorption (solid line). Indicated by a dashed and dotted 
line are the Non-LTE spectrum and the plasma spectrum respectively. Bottom 
panels: The fit of the two-component model with the \sax\ data of \rxj. Note 
that the data above 1.2 keV is no longer in excess of the model}
\label{splitup} 
\end{figure}
The presence of a second spectral component is considered next. The nature of 
this second component is discussed in Sect. \ref{2comp}. It should be kept in 
mind that \rxj\ is only marginally detected in the combined three MECS units, 
i.e. after rebinning only two energy bins between 1.5 and 1.8 keV have a 
significant but low count rate. Even then it is recommended to use only the 
MECS data above $\sim2.0$ keV since the MECS units are poorly calibrated below 
this energy. Since the MECS effective area is approximately three times larger 
than the LECS effective area, the non-detection of \rxj\ above 2.0 keV must 
be used to obtain an upper limit for the luminosity of the second spectral 
component.
\par
We combine the model spectra of GRID2 with a Collisional Ionization Equilibrium 
(CIE) plasma cf. Kaastra et al. (\cite{kaas}) to fit with the data. We make 
several fits to the LECS data with the plasma temperature fixed between 0.2 and 
1.0 keV. For each plasma temperature we find a corresponding emission measure 
(E.M. or $\int n_{\rm e}n_{\rm H}dV$) and a luminosity, see Table 
\ref{2ndcomp}. We 
then simulate a MECS spectrum using the obtained parameters and check whether 
the spectrum would be observed with the MECS. It turns out that at plasma 
temperatures above $\sim 1$ keV the fit with the LECS data becomes increasingly 
worse and the model is even in excess of the data. In addition there are 
several MECS energy bins between 2.0 and 3.0 keV which have significant counts 
above the noise level. For plasma temperatures between 0.2 and 1 keV the 
$\chi^2$ value has decreased to 3.4. From Fig. \ref{splitup} it can be seen 
that this change in $\chi^2$ is mainly due to the disappearance of the excess 
above 1.2 keV. The parameters for the optically thick atmospheric component are 
unaffected by the addition of the CIE spectrum because the photon flux of the 
former is at least a factor 10 larger than that of the latter in the energy 
range in which the atmospheric component is observed (0.2--1.2 keV). It should 
be emphasized that at a plasma temperature of $\la0.6$ keV it is no longer 
possible to distinguish the simulated MECS plasma spectrum from the noise above 
energies of 2.0 keV. This corresponds to a plasma luminosity of $L_{\rm 
CIE}\ga8.8\times 10^{32}\mbox{ erg s}^{-1}\mbox{(d/kpc)}^2$. Note that lower 
plasma temperatures are accompanied by increasingly higher emission measures 
and luminosities. 
\section{Discussion and conclusions} 
\subsection{One component model} 
\label{1comp} 
All single component fits result in a very high value for $\chi^2$. The 
problems are mainly caused by the presence of strong absorption edges in the 
models which are not observed in the \xr\ data. We summarize our findings 
below. 
\par 
Below certain values of the effective temperature (depending e.g. upon the 
surface gravity, see Fig. \ref{jump}) the O\,{\sc viii} edge becomes a 
dominant feature in the model spectrum which is not observed. On the other 
hand, Ebisawa et al. find an indication of a shallow O\,{\sc viii} edge 
(Ebisawa et al. \cite{ebis2}) and thus, cf. Fig. \ref{jump}, there is an upper 
limit to the effective temperature for each value of $\log g$.
\par 
The Ne\,{\sc ix} edge is obviously not observed in the \xr\ spectrum. The 
problem is that adapting the effective temperature such that Ne\,{\sc ix} 
recombines to Ne\,{\sc viii} or ionizes to Ne\,{\sc x} is not compatible with 
our findings for O\,{\sc viii}. A possible solution is to exclude neon from the 
models. This is discussed below. 
\par 
Hot models with $\teff\ga 10^6\mbox{ K}$ develop strong highly ionized iron 
line blends in their spectra. This shows up in the folded spectra as a rather 
flat energy distribution above 0.8 keV. Again this is not observed in the \xr\ 
spectrum. See Fig. \ref{feblend}. One possible way out of this problem is to 
deplete the atmosphere of iron and neon by gravitational settling. This 
mechanism has been discussed by several authors for evolved stars like white 
dwarfs (e.g. Schatzman \cite{schatzman}). The general conclusion is that 
isolated white dwarfs are depleted of heavy elements on timescales short 
compared to their evolutionary time-scales. However, there are several 
arguments against gravitational depletion of metals in a supersoft source.
\par
First, \ssss\ in general are subject to severe accretion from their companion 
star. Dupuis et al. (\cite{dupuis1}, \cite{dupuis2}) have simulated 
gravitational settling of white dwarfs under the influence of accretion from 
the interstellar medium. They assume accretion rates between 
$10^{-20}-10^{-15}\mbox{ M}_{\odot}\mbox{ yr}^{-1}$. At the highest accretion 
rates ($10^{-15}\mbox{ M}_{\odot}\mbox{ yr}^{-1}$) they conclude that metals 
must become observable in white dwarf atmospheres at optical and UV 
wavelengths. For a massive white dwarf $\sim1\mbox{ M}_\odot$ in a \sss\ the 
accretion rate is typically $10^{-7}{\rm M}_\odot \mbox{ yr}^{-1}$ (e.g. Van 
den Heuvel et al. \cite{heuvel}). Note that this is many orders of magnitude 
more than the values used in the computations by Dupuis et al. Thus accreted 
metals should be observable in \ssss.
\par
Second, Chayer et al. (\cite{chayer1}, \cite{chayer2}) have calculated the 
effects for radiative levitation upon the atmospheric abundances. They even 
obtain metal overabundances in the atmospheres of relatively hot white dwarfs.
\par
Third, Prialnik \& Shara (\cite{prial}) have calculated mass accretion, 
diffusion, convection, nuclear burning and hydrodynamic mass loss of massive 
white dwarfs representing classical novae. They find that the oxygen abundance 
is a sensitive function of the white dwarf mass, temperature and accretion rate 
but that the neon abundance is enhanced with respect to solar.\\
Given these three arguments there is no reason to favor metal (neon and iron) 
depletion of the atmosphere of \rxj\ by gravitational settling. Thus we assume 
that the abundances reflect those of the companion star which we consider 
solar. Therefore, we conclude that \rxj\ must have an effective temperature 
below $\teff\approx10^6\mbox{ K}$.
\par
Applying the model spectra of GRID2 the derived luminosity of the optically 
thick component (at 1 kpc) is still an order of magnitude too low in order to 
`lift' it into the shaded region in Fig. \ref{lt_plot}. The low inclination of 
the system (Motch \cite{motch3}) rules out the possibility of obscuration of 
the white dwarf by the accretion disk. We therefore conclude that the derived 
luminosity of \rxj\ is consistent with that of a white dwarf located at a 
distance of $\sim4\mbox{ kpc}$.
\subsection{Two component model} 
\label{2comp} 
The best fit result is obtained when we introduce hot plasma emission as a 
second spectral component next to the atmospheric emission. With this model it 
is not necessary to assume low metal (neon) abundances. The fit parameters for 
the atmospheric emission do not change when a second component is introduced 
since the bulk of the photon flux between 0.1--1.2 keV still arises from the 
model atmosphere. The presence of a plasma component next to optically thick 
atmospheric emission has been investigated by Balman et al. (\cite{balman}) in 
the case of ROSAT observations of the classical nova V1974 Cygni. They find 
that the relatively soft (optically thick) spectrum fits considerably better 
when they include a second, optically thin Raymond-Smith plasma at temperatures 
$T_{\rm RS}\la 1\mbox{ keV}$. Their explanation is that a fast wind from the 
nova collides with material ejected earlier from the system. A similar model 
for the optically thin emission in \rxj\ may be adopted since there are clear 
evidences for P Cygni profiles at ${\rm H}\alpha$ which indicate a (constant) 
bulk wind from the system (Motch \cite{motch2}). The mass-loss rate of a wind 
with constant velocity outflow can be expressed by: 
\bea
\dot{M} \approx 2v_{\infty}\,m_{\rm H} \left({EM \over N_{\rm H}}\right),
\nonumber
\eea
(cf. Balman et al \cite{balman}) where $v_{\infty}$ is the escape velocity and 
$N_{\rm H}$ is the intrinsic column density due to the mass outflow. A simple 
estimate using $\dot{M}\approx 10^{-7}\mbox{ M}_\odot\mbox{yr}^{-1}$, $v_\infty 
\approx 10^8\mbox{ cm s}^{-1}$ and $N_{\rm H} \la 10^{22}\mbox{ cm}^{-2}$ shows 
that the expected emission measure is of the order of $\sim 10^{56}\mbox{ 
cm}^{-3}$, consistent with the emission measure we derive from the 
two-component fit (see Table \ref{2ndcomp}). Balman et al. observe similar 
values for the emission measure in the case of V1974 Cygni.
\par
We assume that the wind originates from a white dwarf and compare the 
kinetic energy content of such a wind with the plasma luminosity of $L_{\rm 
CIE}\ga8.8\times 10^{32}\mbox{ erg s}^{-1}\mbox{(d/kpc)}^2$ suggested in Sect. 
\ref{comp_solution}. The energy carried away by the wind is given by:
\bea
\dot{E}_{\rm wind}={GM_{\rm WD} \over R_{\rm WD}}\,\dot{M}_{\rm wind}.
\nonumber
\eea
Such a wind can collide with a shell ejected during a helium flash occuring at 
an earlier evolutionary phase when the nuclear burning is not stable. If we 
assume that a fraction $\phi$ of the kinetic energy of the wind is radiated 
away at the shock then
\bea
\phi\dot{E}_{\rm wind}\approx L_{\rm CIE}\ga8.8\times 10^{32}\mbox{ erg 
s}^{-1}\mbox{(d/kpc)}^2.
\nonumber
\eea
Taking $M_{\rm WD}\approx 1.1\mbox{ M}_\odot$ and $R_{\rm WD}\approx 4300\mbox{ 
km}$ (corresponding roughly to $\log g = 9.0$) then
\bea
\dot{M}_{\rm wind}\ga 4 \times 10^{-11}\phi^{-1} \mbox{M}_\odot \mbox{yr}^{-1} 
\mbox{(d/kpc)}^2,
\nonumber
\eea
and
\bea
\dot{M}_{\rm wind}\la\dot{M}_{\rm 
WD}\approx10^{-7}\mbox{M}_\odot\mbox{yr}^{-1}.
\nonumber
\eea
The latter criterion comes from the fact that the wind mass loss can not exceed 
the mass accretion from the donor star for stable nuclear burning white dwarfs. 
Even when \rxj\ is located at a distance of 4 kpc (see Sect. \ref{1comp}) the 
efficiency factor $\phi$ can be less than a percent in order to make the 
observed plasma luminosity consistent with a shocked fast wind.
\par
Van Teeseling \& King (\cite{tees}) have calculated the mass loss rate from the 
companion star in a \sss\ in general as a result of irradiation by the white 
dwarf. They find that the companion star loses mass as a result of the 
irradiation with a rate of $10^{-7}-10^{-6}\mbox{ M}_\odot\mbox{yr}^{-1}$. 
Assuming a typical wind velocity of $\sim100\mbox{ km s}^{-1}$ we obtain for 
the energy carried away by the wind
\bea
\dot{E}_{\rm wind}={1 \over 2}\,\dot{M}_{\rm 
wind}v^2\approx3\times 10^{33}\mbox{ erg s}^{-1}.
\nonumber
\eea
Obviously this value is too low to account for the observed plasma luminosity, 
even if we use a shock efficiency $\phi=1$. However, this comparison is 
sensitive to the assumed values for the wind velocity and the distance to \rxj. 
Note that only a small fraction of the optically thick luminosity needs to be 
reprocessed in the atmosphere of the companion star in order to account for the 
plasma luminosity. We consider optically thin emission from a shocked wind or 
from the irradiation of the companion star the best options for the explanation 
of the observed flux above 1.2 keV.\\

The analysis of the \sax\ LECS spectrum of \rxj\ has been severely restricted 
due to the moderate spectral resolution and rebinning of the energy channels 
because of the low count rate. However, \rxj\ has been scheduled for 
observation with the Advanced \xr\ Astrophysics Facility (AXAF). High 
resolution spectroscopy with $E/\Delta E$ of the order of several hundred, 
particularly at low energies, combined with a reduced background per beam 
element will allow to study the energy distribution of this source in much more 
detail. We expect to be able to refine our statements about the nature of the 
emission above $\sim1\mbox{ keV}$. Especially when the emission originates from 
an optically thin plasma metal emission lines may be observed.
\begin{acknowledgements}
This work has been supported by funds of the Netherlands Organization for 
Scientific Research (NWO). 
\end{acknowledgements} 
\end{document}